\documentclass[12pt,a4paper]{article}
\begin{document}
\large

\newpage
\begin{center}
{\bf THE UNITED FORCES IN THE NATURE OF MATTER}
\end{center}
\vspace{1cm}
\begin{center}
{\bf Rasulkhozha S. Sharafiddinov}
\end{center}
\vspace{1cm}
\begin{center}
{\bf Institute of Nuclear Physics, Uzbekistan Academy of Sciences,
Tashkent, 702132 Ulugbek, Uzbekistan}
\end{center}
\vspace{1cm}

The united rest mass and charge of a particle correspond to two form of the
same regularity of the unified nature of its ultimate structure. Each of them
contains the electric, weak, strong and the gravitational contributions. As a
consequence, the force of an attraction among two neutrinos and force of their
repulsion must be defined from point of view of any of existing types of the
actions. Therefore, to understand the nature of the micro world interaction at
the fundamental level one must use the fact that each of the four well known
forces includes in self both a kind of the Newton and a kind of the Coulomb
components. The opinion has been speaked out that the existence of the
gravitational parts of the united rest mass and charge implies the
availability of the fifth force in the nature of matter.

\newpage
In studying the fundamental structure of matter such characteristics
as the mass and charge play a large role. At the same time their nature
remains thus far not finally established. Usually it is accepted that
there is no any connection between the mass and charge. However, according
to the hypothesis of field mass based on the classical model of a hard
electron \cite{1}, a particle all the mass is strictly electric.

Our analysis of the behavior of massive neutrinos in a nucleus field
shows \cite{2,3} clearly that between the mass of a Dirac neutrino and
its electric charge there exists an intimate interconnection. Such a sharp
dependence giving the possibility to investigate the compound structure of
charge quantization law \cite{4} and reflects the fact that each of existing
types of charges testifies in favor of the availability of a kind of the
inertial mass \cite{5}. Thereby this mass - charge duality of matter explains
the coexistence of the united rest mass and charge for the same particle.
At the account of earlier findings, nonweak \cite{6,7} and undiscovered
properties of the neutrino they have the form \cite{5}
\begin{equation}
m_{\nu}^{U}=m_{\nu}^{E}+m_{\nu}^{W}+m_{\nu}^{S}+...,
\label{1}
\end{equation}
\begin{equation}
e_{\nu}^{U}=e_{\nu}^{E}+e_{\nu}^{W}+e_{\nu}^{S}+....
\label{2}
\end{equation}
Here the indices $E,$ $W$ and $S$ imply that both mass and charge
of the neutrino contain as well as the electric, weak and strong components.

Exactly the same one can as the development of these sights include in the
discussion the gravitational mass and charge. This procedure, however, meets
with many problems. One of them states that the united rest mass of a particle
contains the part that corresponds to its gravitational charge.

In conformity with such contributions of the gravitational mass $m_{\nu}^{G}$
and charge $e_{\nu}^{G},$ we not only recognize \cite{8} that
\begin{equation}
m_{\nu}^{U}=m_{\nu}^{E}+m_{\nu}^{W}+m_{\nu}^{S}+m_{\nu}^{G},
\label{3}
\end{equation}
\begin{equation}
e_{\nu}^{U}=e_{\nu}^{E}+e_{\nu}^{W}+e_{\nu}^{S}+e_{\nu}^{G}
\label{4}
\end{equation}
but also need elucidate what neutrino united rest mass and charge say about
the unified force in the nature of matter. The answer to this question one
can obtain by studying the ideas of each of existing types of charges and
masses. All they therefore will be illuminated in the present work.

According to our presentations about the structural properties of rest mass
and charge, the force of gravity of the Newton $F_{N_{\nu\nu}}$ among two
neutrinos and force of the Coulomb $F_{C_{\nu\nu}}$ between themselves may
also be expressed from point of view of any of all possible types of the
actions. In other words, each of these forces becomes the function as well
as of corresponding components of the united masses or charges of interacting
objects.

One can define their structure in the limit of the electric masses and
charges as follows \cite{5}:
\begin{equation}
F^{E}_{N_{\nu\nu}}=
G\left(\frac{m_{\nu}^{E}}{R}\right)^{2}, \, \, \, \,
F^{E}_{C_{\nu\nu}}=\frac{1}{4\pi\epsilon_{0}}
\left(\frac{e_{\nu}^{E}}{R}\right)^{2},
\label{5}
\end{equation}
where $G$ is the constant of the gravitational action.

These forces for the weak masses and charges may have the form
\begin{equation}
F^{W}_{N_{\nu\nu}}=
G\left(\frac{m_{\nu}^{W}}{R}\right)^{2}, \, \, \, \,
F^{W}_{C_{\nu\nu}}=\frac{1}{4\pi\epsilon_{0}}
\left(\frac{e_{\nu}^{W}}{R}\right)^{2}.
\label{6}
\end{equation}

At the availability of strong masses and charges they must behave as
\begin{equation}
F^{S}_{N_{\nu\nu}}=
G\left(\frac{m_{\nu}^{S}}{R}\right)^{2}, \, \, \, \,
F^{S}_{C_{\nu\nu}}=\frac{1}{4\pi\epsilon_{0}}
\left(\frac{e_{\nu}^{S}}{R}\right)^{2}.
\label{7}
\end{equation}

The contributions explained by the gravitational masses and charges
of interacting particles are written in the form
\begin{equation}
F^{G}_{N_{\nu\nu}}=
G\left(\frac{m_{\nu}^{G}}{R}\right)^{2}, \, \, \, \,
F^{G}_{C_{\nu\nu}}=\frac{1}{4\pi\epsilon_{0}}
\left(\frac{e_{\nu}^{G}}{R}\right)^{2}.
\label{8}
\end{equation}

In a similar way the studyied forces may also be defined for the united
masses and charges:
\begin{equation}
F^{U}_{N_{\nu\nu}}=
G\left(\frac{m_{\nu}^{U}}{R}\right)^{2}, \, \, \, \,
F^{U}_{C_{\nu\nu}}=\frac{1}{4\pi\epsilon_{0}}
\left(\frac{e_{\nu}^{U}}{R}\right)^{2}.
\label{9}
\end{equation}

Comparison of (\ref{9}) with the corresponding size from (\ref{5}) - (\ref{8})
leads us to the consequence of correspondence principle that
\begin{equation}
F^{U}_{N_{\nu\nu}}=F^{E}_{N_{\nu\nu}}+F^{W}_{N_{\nu\nu}}+
F^{S}_{N_{\nu\nu}}+F^{G}_{N_{\nu\nu}},
\label{10}
\end{equation}
\begin{equation}
F^{U}_{C_{\nu\nu}}=F^{E}_{C_{\nu\nu}}+F^{W}_{C_{\nu\nu}}+
F^{S}_{C_{\nu\nu}}+F^{G}_{C_{\nu\nu}}.
\label{11}
\end{equation}

Insertion of (\ref{3}) and (\ref{4}) in (\ref{9}) it would seem says
of that equations (\ref{10}) and (\ref{11}) do not correspond to the reality
at all. It is easy to observe, however, that this is not quite so. The point
is that any particle with the united mass and charge come forwards in the
system as the unified and the whole. Nobody is in force to separate its
by part in the mass or charge type dependence.

Furthermore, if it turns out that just the compound structures of the united
rest mass and charge establish the intraneutrino harmony of forces of the
different nature \cite{9}, this can also confirm the fact that $m_{\nu}^{U}$
and $e_{\nu}^{U}$ are the multicomponent vectors, squares of which
become equal to
\[|\vec{m_{\nu}^{U}}|^{2}=
(m_{\nu}^{E})^{2}+(m_{\nu}^{W})^{2}+(m_{\nu}^{S})^{2}+(m_{\nu}^{G})^{2},\]
\[|\vec{e_{\nu}^{U}}|^{2}=
(e_{\nu}^{E})^{2}+(e_{\nu}^{W})^{2}+(e_{\nu}^{S})^{2}+(e_{\nu}^{G})^{2}\]
and that, consequently, the existence of solutions (\ref{10}) and (\ref{11})
is by no means excluded naturally.

The absence of one of forces $F_{N_{\nu\nu}}$ or $F_{C_{\nu\nu}}$ would
imply that both do not exist at all \cite{5}. This becomes possible owing
to the mass - charge duality of matter. In other words, $F_{N_{\nu\nu}}$
and $F_{C_{\nu\nu}}$ correspond to the most diverse form of the same action
at the different distances.

According to the recent presentations about the nature of strong matter,
the nuclear forces at a small distances have the character of the repulsion.
In a large distances dependence can essentially appear their property of an
attraction. Taking into account these facts and all what neutrino masses
and charges say about structures of fundamental forces, we are led to the
implication that each of them includes in self those parts which correspond
to the masses and charges of interacting objects. Formulating more concretely,
one can write the electric $F^{E}_{\nu\nu},$ weak $F^{W}_{\nu\nu},$ strong
$F^{S}_{\nu\nu}$ and the gravitational $F^{G}_{\nu\nu}$ forces of the
interaction between the particles in general form:
\begin{equation}
F^{E}_{\nu\nu}=F^{E}_{N_{\nu\nu}}+F^{E}_{C_{\nu\nu}},
\label{12}
\end{equation}
\begin{equation}
F^{W}_{\nu\nu}=F^{W}_{N_{\nu\nu}}+F^{W}_{C_{\nu\nu}},
\label{13}
\end{equation}
\begin{equation}
F^{S}_{\nu\nu}=F^{S}_{N_{\nu\nu}}+F^{S}_{C_{\nu\nu}},
\label{14}
\end{equation}
\begin{equation}
F^{G}_{\nu\nu}=F^{G}_{N_{\nu\nu}}+F^{G}_{C_{\nu\nu}}.
\label{15}
\end{equation}

So, it is seen that any of the four well known forces contains both a kind
of the Newton and a kind of the Coulomb parts. In this the unified regularity
is said of the nature of these forces. Therefore, to create at the fundamental
level a truly picture of the micro world interaction one must establish the
compound structures of their naturally united gauge potentials.

If now the idea of any of forces (\ref{9}) is accepted, it should
be added that
\begin{equation}
F^{U}_{\nu\nu}=F^{U}_{N_{\nu\nu}}+F^{U}_{C_{\nu\nu}}.
\label{16}
\end{equation}

Using (\ref{10}), (\ref{11}) and taking (\ref{12}) - (\ref{15}), we
find that
\begin{equation}
F^{U}_{\nu\nu}=F^{E}_{\nu\nu}+F^{W}_{\nu\nu}+
F^{S}_{\nu\nu}+F^{G}_{\nu\nu}.
\label{17}
\end{equation}

Thus, it follows that if the gravitational rest mass and charge of a
particle are unequal to its all the mass and charge, this will indicate
to the existence of the fifth force of the interaction which come forwards
in the nature as the unified and the united.

\end{document}